\documentclass[12pt]{iopart}

\usepackage{iopams}  
\usepackage{graphicx}

\begin{document}

\title{ In-plane thermal conductivity of large single crystals of  Sm-substituted  (Y$_{1-x}$Sm$_{x}$)Ba$_{2}$Cu$_{3}$O$_{7-\delta}$}

\author{ M.Matsukawa,H.Noto and H.Furusawa}
\ead{matsukawa@iwate-u.ac.jp}
\address{ Department of Materials Science and Technology, Iwate University , Morioka 020-8551 , Japan }
\author{ X.Yao}

\address{Department of Physics, Shanhai Jiao Tong University, Shanghai 200030, People's Republic of China}

\author{ S.Nimori }

\address{ National Institute for Materials Science, Tsukuba 305-0047 ,Japan }
\author{ N. Kobayashi }
\address{ Institute for Materials Research, Tohoku University, Sendai 980-8577, Japan }
\author{ Y.Shiohara}
\address{ Superconducting Research Laboratory, ISTEC,Tokyo 135-0062, Japan }

\begin{abstract}
We have investigated  the  in-plane thermal conductivity  $\kappa _{ab}(T,H)$ of large single crystals of optimally oxygen-doped
 (Y$_{1-x}$,Sm$_{x}$)Ba$_{2}$Cu$_{3}$O$_{7-\delta}$ ($x$=0, 0.1, 0.2 and 1.0) and YBa$_{2}$(Cu$_{1-y}$Zn$_{y}$)$_{3}$O$_{7-\delta}$( $y$=0.0071)
as functions of temperature and magnetic field (along the $c$ axis).  For comparison,  
the temperature dependence of  $\kappa _{ab}$ for as-grown crystals with the corresponding compositions are presented.
 The nonlinear field dependence of  $\kappa _{ab}$ for all crystals  was observed at relatively low fields near a half of $T_{c}$. 
We make fits  of the $\kappa(H)$  data to an electron contribution model, providing both the mean free path of quasiparticles  $\ell_{0}$ and the electronic thermal conductivity $\kappa _{e}$, in the absence of field.  
The local lattice distortion due to the Sm substitution for Y  suppresses both the phonon and electron contributions. 
On the other hand, the light Zn doping into the CuO $_{2}$ planes affects solely the electron component below $T_{c}$, resulting in a substantial decrease in  $\ell_{0}$ .
%For  SmBa$_{2}$Cu$_{3}$O$_{7-\delta}$, a suppressed maximum in  $\kappa(T)$ arises from the partial substitution of Sm for Ba which is associated with the slight degradation in $T_{c}$. 
\end{abstract}

\maketitle

\section{Introduction}

Since the discovery of high-$T_{c}$ cuprate with its perovskite structure,  a large number of measurements of thermal conductivity  $\kappa$ have been reported up to date, giving crucial information about transport properties in the superconducting and normal states\cite{UH90}. 
The thermal conductivity of solids is usually separated into the phonon and electronic components $\kappa_{ph}$ and $\kappa_{e}$ 
\cite{ZI60}.  
 For  high-$T_{c}$ copper oxide superconductors YBa$_{2}$Cu$_{3}$O$_{7-\delta}$(YBCO), 
Bi$_{2}$Sr$_{2}$CaCu$_{2}$O$_{8+\delta}$ and Tl$_{2}$Sr$_{2}$CaCu$_{2}$O$_{8+\delta}$ , a rapid enhancement in $\kappa(T)$  is observed below $T_{c}$, with decreasing  $T$, then $\kappa(T)$ reaches a maximum around a half of $T_{c}$ and finally shows a monotonous decrease  at low temperatures. 
Tewordt et al. ascribed the origin of the anomaly to the phonon contribution $\kappa_{ph}$ on the basis of the classical theory of thermal transport in superconductors \cite{TW89}. On the contrary, Yu et al. have proposed  a new interpretation of the anomalous enhancement in $\kappa(T)$ as being responsible for the electron contribution $\kappa_{e}$ \cite{YU92}. Following their proposal, a strong suppression  in the quasiparticle scattering rate below $T_{c}$ causes the  $\kappa(T)$ anomaly, which is consistent with  an observed peak in measurements of microwave conductivity in  YBa$_{2}$Cu$_{3}$O$_{7-\delta}$ \cite{BO93}.
In addition, the magnetic field effect on the thermal conductivity in high-$T_{c}$ superconductors enables us to examine the scattering processes of thermal carriers due to magnetic vortex lines \cite{PE91,RI91}.
In ref. \cite{YU94}, the authors  developed a two component method to analyze the field dependence of the thermal conductivity and concluded that only the electronic contribution is field sensitive. In a similar way, a procedure to check the linear field dependence of inverse electronic thermal conductivity was proposed by Pogorelov et al. \cite{PO95}.
%Recent studies on thermal Hall conductivity on single crystals 
%YBa$_{2}$Cu$_{3}$O$_{7-\delta}$ support the electronic origin of the $\kappa$ peak 
%\cite{KR95,KR99}.  
However, the origin of the anomalous peak in $\kappa(T)$ and its field dependence are still open questions.

In this paper, we present measurements of the in-plane thermal conductivity $\kappa _{ab}$ of  Sm-substituted single crystals (Y$_{1-x}$,Sm$_{x}$)Ba$_{2}$Cu$_{3}$O$_{7-\delta}$ ($x$=0, 0.1, 0.2 and 1.0) as 
functions of temperature and magnetic field. To our knowledge, no measurement of thermal conductivity has been performed on A-site substituted single crystals YBCO except for polycrystalline samples \cite{TI94,WI99}. Here, we give some comments on the substitution effect of Sm ion on the crystal growth of YBa$_{2}$Cu$_{3}$O$_{7-\delta}$ . In a previous work, the substitution of Sm for Y led to a high growth rate for single crystal growth, which is about two times greater with a Sm content $x$=0.12 than that of YBCO  \cite{YA96}.  
Moreover, by controlling the Sm content, one obtained high superconductivity for Sm substituted YBCO. 

 The $\kappa(H)$ profile is fitted using the electron contribution model, to estimate the mean free path of quasi particles  $\ell_{0}$ near a half of $T_{c}$ in the absence of field \cite{YU94,KR95}.  Compared with the linear field dependence of the inverse $\kappa_{e}$ ,the square-root fit to the  $\kappa(H)$data  is applied .  
In addition, we discuss the impurity substitution effect on the thermal conduction of  large single crystals YBa$_{2}$Cu$_{3}$O$_{7-\delta}$. 

\section{Experiment}

 A series of single crystalline
 (Y$_{1-x}$,Sm$_{x}$)Ba$_{2}$Cu$_{3}$O$_{7-\delta}$ ($x$=0, 0.1, 0.2 and 1.0) and YBa$_{2}$(Cu$_{1-y}$Zn$_{y}$)$_{3}$O$_{7-\delta}$ ($y$=0.0071)  were grown by the crystal pulling method.
Sm contents range from 0, 0.1(0.119),0.2(0.215) and 1.0. As-grown crystals with $x$=0, 0.1, 0.2 and $y$(Zn)=0.0071 were annealed at 
$520^{\circ}$C  in the oxygen atmosphere (for 1 to 2 weeks)to achieve the optimal $T_{c}$ with $7-\delta=\sim 6.9$, as shown in Fig.\ref{MT}.  SmBa$_{2}$Cu$_{3}$O$_{7-\delta}$ (SmBCO) crystal was oxidized  at $300^{\circ}$C for 200h, giving a higher $T_{c}$ property. 
The slight degradation in $T_{c}$ for SmBCO probably arises from the partial substitution of Sm for Ba site rather than
oxygen defects because we  annealed it in the oxygen atmosphere for a very long time. 
The oxygen content for the as-grown crystal with $x$=0 was estimated to be $7-\delta=\sim 6.17$ using the iodometric analysis on one piece cut from the same batch. 
Sample preparations are described by Yao et al.in 
\cite{YA96,YA03,YAZ96}. 
Sample dimensions are typically 4.5$\times $3.1 mm$^2$ in the $ab$-plane and 1.7 mm along the $c$-axis.
The EPMA analysis showed that Sm ion  is uniformly distributed in single crystals. The inductively coupled plasma (ICP) technique revealed that single crystalline samples used in our measurements are close to the nominal compositions. 

Measurements of the thermal conductivity, in the $ab$-plane and along the $c$-axis, were done with a conventional steady-state heat-flow method\cite{MA96}. The magnetic field effect on the $ab$-plane thermal conductivity was examined 
in the application of fields up to 14 T along the $c$-axis 
at the Tsukuba Magnet Laboratory, the National Institute for Materials Science and at the High Field Laboratory for 
Superconducting Materials, Institute for Materials Research, Tohoku University.
A temperature gradient across the sample  was determined using a differential-type  Chromel-constantan thermocouple after a   calibration procedure.   The magnetization measurement was carried out using the SQUID magnetometer.  The temperature variation of the resistivity  was measured on the identical samples using a four-probe technique.   
 
\section{Results and discussion}

\begin{figure}[p]
\includegraphics[width=12cm]{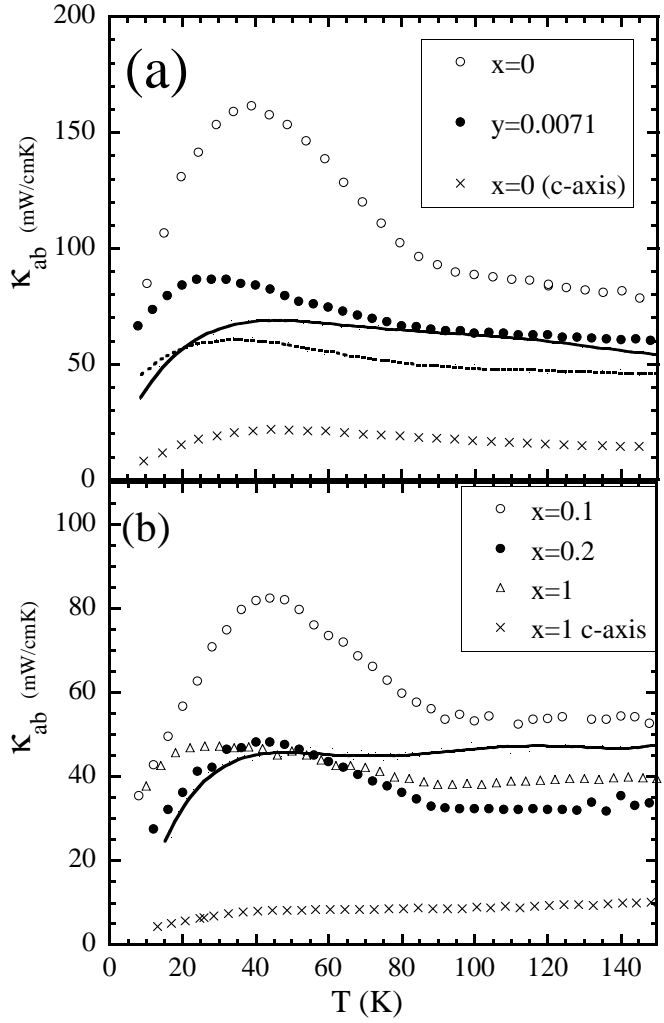}%
\caption{ The in-plane thermal conductivity $\kappa _{ab}$ for single crystals  (Y$_{1-x}$,Sm$_{x}$)Ba$_{2}$Cu$_{3}$O$_{7-\delta}$ ($x$=0, 0.1, 0.2 and 1.0) and YBa$_{2}$(Cu$_{1-y}$Zn$_{y}$)$_{3}$O$_{7-\delta}$ ($y=0.0071$)  as a function of temperature.  For comparison, the out-of-plane thermal conductivity $\kappa _{c}$with $x$=0 and 1.0 is also depicted. 
Solid and dashed curves in (a) represent  the estimated phonon component $\kappa _{ph}$  for $x$=0 and $y=0.0071$,respectively (see to the text). In a similar way, a solid curve in (b) shows the phonon contribution $\kappa _{ph}$ of the $x$=0.1 crystal. }
\label{KT}
\end{figure}

\begin{figure}[p]
\includegraphics[width=12cm]{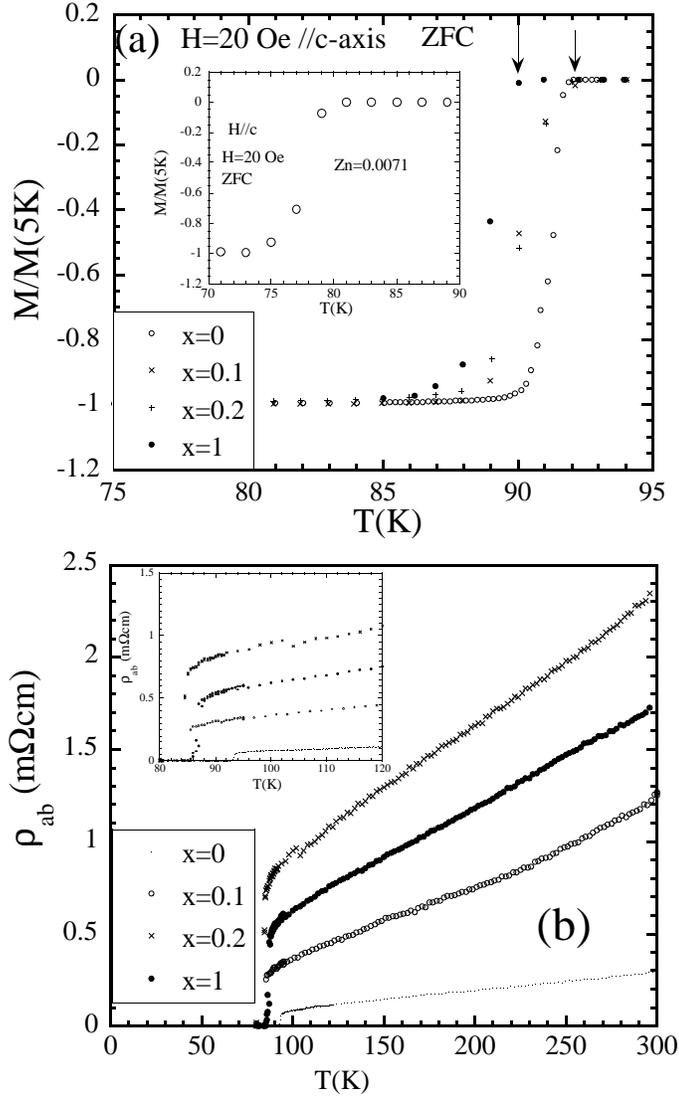}%
\caption{(a)normalized low-field magnetization $M(T)$ under a zero-field cooling (ZFC) scan and (b) in-plane resistivity $\rho _{ab}(T)$, for single crystals  (Y$_{1-x}$,Sm$_{x}$)Ba$_{2}$Cu$_{3}$O$_{7-\delta}$ ($x$=0, 0.1, 0.2 and 1.0).
The $M(T)$ data of YBa$_{2}$(Cu$_{1-y}$Zn$_{y}$)$_{3}$O$_{7-\delta}$ ($y=0.0071$) are given in the inset of (a).  The inset of (b) represents magnified plots for the $\rho _{ab}(T)$ data of the Sm-substituted samples.  }
\label{MT}
\end{figure}

The in-plane thermal conductivity $\kappa _{ab}$ for single crystals   (Y$_{1-x}$,Sm$_{x}$)Ba$_{2}$Cu$_{3}$O$_{7-\delta}$ ($x$=0, 0.1, 0.2 and 1.0) and YBa$_{2}$(Cu$_{1-y}$Zn$_{y}$)$_{3}$O$_{7-\delta}$ (y=0.0071) is shown as a function of temperature in Fig.\ref{KT}.  For comparison, the out-of-plane thermal conductivity $\kappa _{c}$with $x$=0 and 1.0 is also depicted in the Fig.\ref{KT}. 
 For the identical crystals used for our experiment , measurements of the low-field magnetization and the in-plane resistivity are presented in Fig.\ref{MT}, to check the superconducting properties. The resistivity data for the lightly Zn doped sample will be shown in the inset of Fig.3(b).

First of all, the $\kappa _{ab}$ of (Y$_{1-x}$,Sm$_{x}$)Ba$_{2}$Cu$_{3}$O$_{7-\delta}$ showed a rapid enhancement upon crossing $T_{c}$ with decreasing temperature,  while for   $\kappa _{c}$  no remarkable variation was detected near $T_{c}$.  The substitution of Y  by Sm strongly suppresses a maximum of $\kappa_{ab}$ from 160 mW/cmK at $x$=0 down to 50 mW/cmK at $x$=0.2  through 85 mW/cmK at $x$=0.1. However, the pronounced peak of $\kappa _{ab}$ below $T_{c}$ is still observed even for the $x$=0.2 crystal with its low thermal value. 
For the SmBCO crystal,  $\kappa _{ab}$ exhibits a low thermal conductivity accompanied by the broad maximum near 40K.  This finding is closely related to a local lattice distortion due to the slight substitution of Sm for Ba  as mentioned before.
For free ions, the effective moment of  Sm$^{3+}$ is 0.85 $\mu_{B}$ and it is much smaller than the value of other rare earth ions (3.62 $\mu_{B}$ for Nd$^{3+}$ and 7.96 $\mu_{B}$ for Gd$^{3+}$).  If the phonon scattering due to magnetic ions with the smaller moment  is assumed to be in the presence of the crystal field for  SmBCO, it will contribute  a suppression observed in  $\kappa _{ab}$ to some extent.   

\begin{figure}[p]
\includegraphics[width=12cm]{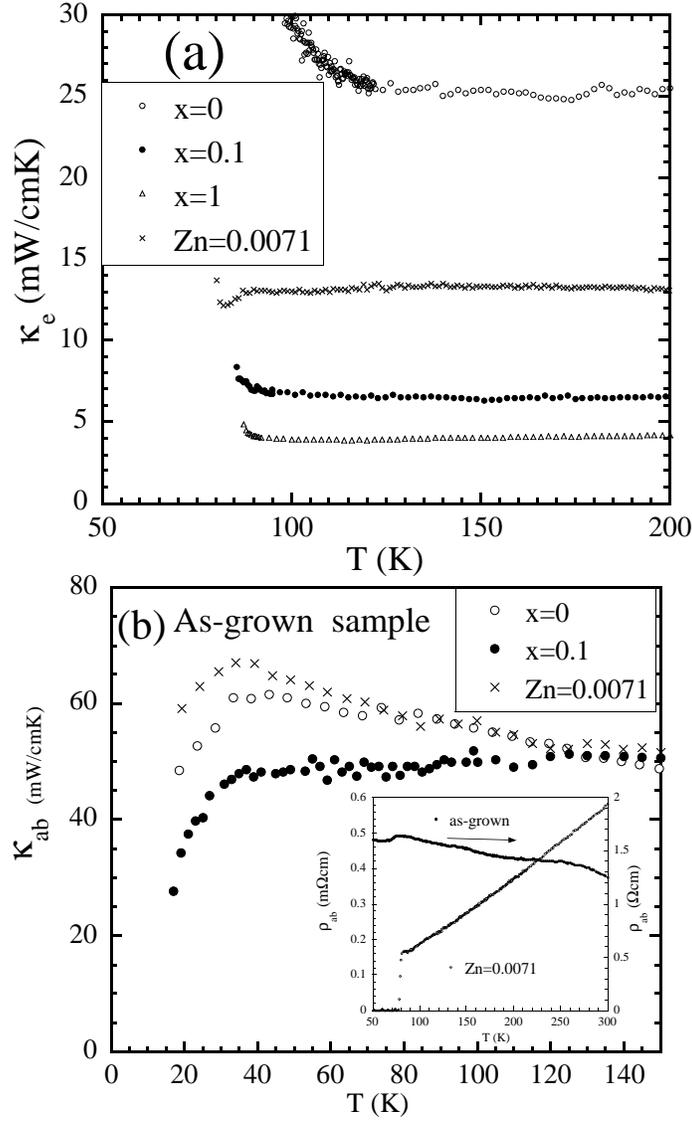}%
\caption{(a) Normal-state electronic thermal conductivity $\kappa_{e}^{n}$ estimated from the resistivity data using the Wiedemann-Franz (WF) law $\kappa_{e}^{n}=L_{0}T/\rho $ with a Lorentz number $L_{0}=$2.45$\times 10^{-8}$W$\Omega /$K$^{2}$. (b) The $\kappa _{ab}$ data for the corresponding as-grown crystals of (Y$_{1-x}$,Sm$_{x}$)Ba$_{2}$Cu$_{3}$O$_{7-\delta}$ ($x$=0 and 0.1) and YBa$_{2}$(Cu$_{1-y}$Zn$_{y}$)$_{3}$O$_{7-\delta}$ ($y$=0.0071).
In the inset of (b), the $\rho _{ab}(T)$ data for the as-grown and oxidized samples with light Zn doping are presented.}
\label{KE}
\end{figure}

Now, we comment on $\kappa$ in the normal state for the superconducting crystals.
The thermal conductivity is separated into the electronic and phonon components.
First, the electronic thermal conductivity in the normal state  $\kappa_{e}^{n}$ is estimated from the resistivity data using the Wiedemann-Franz (WF) law $\kappa_{e}^{n}=L_{0}T/\rho $ with a Lorentz number $L_{0}=$2.45$\times 10^{-8}$W$\Omega /$K$^{2}$. For all superconducting crystals, the estimated value  $\kappa_{e}^{n}$ is almost independent of $T$ because of the $T$ linear dependence of $\rho $, as shown in Fig.\ref{KE}(a). 
Accordingly, when $\kappa _{ab}-\kappa_{e}^{n}$ gives rise to the phonon component $\kappa_{ph}(T)$, the $T$ dependence of $\kappa _{ab}(T)$ observed represents that of the phonon component in the normal state \cite{TA97}.

For comparison, let us show in Fig.\ref{KE}(b) the $\kappa _{ab}$ data for the corresponding as-grown crystals of (Y$_{1-x}$,Sm$_{x}$)Ba$_{2}$Cu$_{3}$O$_{7-\delta}$ ($x$=0 and 0.1) and YBa$_{2}$(Cu$_{1-y}$Zn$_{y}$)$_{3}$O$_{7-\delta}$ ($y$=0.0071) .
In the inset of Fig.3(b), the as-grown $\rho $ data for the Zn-doped sample are presented and the electronic component $\kappa_{e}^{n}$ is neglected  from the WF law using its resistivity data.  
For other as-grown crystals, we also expect an insulating  property or a weakly metallic one with  high resistivity.  We note that the as-grown crystals for our experiment were prepared  at least under the similar crystal growth environment. Accordingly,  we conclude that for all as-grown crystals used, $\kappa \approx \kappa_{ph}$. 

The $\kappa _{ab}$ at $x$=0  follows a weakly $T^{-1}$-like behavior,  while for the Sm - substituted crystal it is almost independent on $T$. This finding is probably caused by the enhanced phonon scattering due to the Sm substitution.
Here, At high temperatures, the $T$ dependence of  $\kappa _{ab}(T)$ of the as-grown crystal is expressed as $\kappa_{ ph} = 1/(W_{0}+\alpha T)$,
where $W_{0}$ and $\alpha $  are thermal resistances due to defect scattering and phonon-phonon Umklapp scattering, respectively.
Moreover, for the light Zn doping for the as-grown sample we observed a similar $T$ dependence of $\kappa$ as that in the pure sample, indicating that a small amount of Zn-impurity does not affect the phonon thermal conduction.   
In contrast, the light Zn doping for the superconducting crystal strongly suppresses $T_{c}$  down to 80K  through the superconducting pair breaking, resulting in a reduced peak in $\kappa _{ab}(T)$ in Fig.\ref{KT}(a)\cite{HI96}.

We notice a common trend in the temperature dependence of phonon component  between as-grown and superconducting samples. For example,  $\kappa _{ab}(T)$ of both as-grown and oxygen-annealed samples with $x$=0 shows  a weakly $T^{-1}$-like behavior at high-$T$. 
In addition,  the oxygen content does not affect the $T$ dependence of $\kappa_{ph}$ strongly except for highly insulating crystal YBa$_{2}$Cu$_{3}$O$_{6+\delta}$($\delta \leq 0.11$) \cite{CO95,TA97}.   
Accordingly,  let us  assume that a minor modification of  $\kappa _{ab}(T)$ of as-grown crystals expresses  
the temperature variation of the phonon component below $T_{c}$ for superconducting crystals
if we ascribe the $\kappa$ peak to the electronic contribution. 
Thus, the  original $\kappa$ data for as-grown crystals are scaled  to coincide with $\kappa _{ab}-\kappa_{e}^{n}$  for the corresponding superconducting crystals at high temperatures (150K). 
The solid and dashed curves in Fig.\ref{KT}\ represent the phonon contribution and , at  the selected temperature of 40K, we obtain   43 $\%$ at $x=$ 0,  55  $\%$ at $ x=$ 0.1 and  71 $\%$ at $y$(Zn)=0.0071. (the  corresponding electronic component is listed in Table 1) 
This rough estimation in  $\kappa_{ph}$ is comparable with the value ($\sim40\%$ near $T_{c}$/2 for YBa$_{2}$Cu$_{3}$O$_{7-\delta}$ with $7-\delta=6.93$  ) evaluated by Takenaka et al., \cite{TA97}. 
Separately, a primitive estimation on the phonon component of an untwinned crystal YBCO with $T_{c}$=90.5K  gives a similar value \cite{YU92}. 

\begin{figure}[p]
\includegraphics[width=12cm]{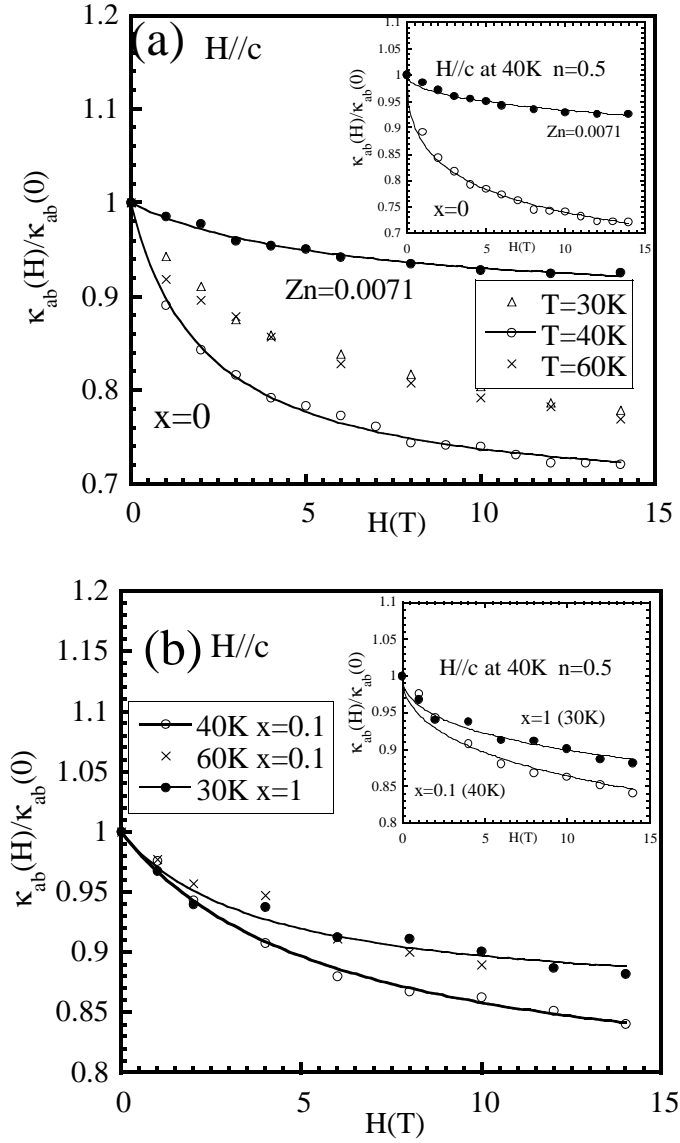}%
\caption{ The magnetic field dependence of the in-plane thermal conductivity 
$\kappa _{ab}(H)$ of single crystals
  (Y$_{1-x}$,Sm$_{x}$)Ba$_{2}$Cu$_{3}$O$_{7-\delta}$ ($x$=0, 0.1 and 1.0) and YBa$_{2}$(Cu$_{1-y}$Zn$_{y}$)$_{3}$O$_{7-\delta}$ ( $y$=0.0071), at selected temperatures.  The applied field $H$ is parallel to the $c$-axis.
Solid curves represent better fits of the $\kappa(H)$ data to the $H$-linear dependence of the inverse
electronic thermal conductivity in Eq.(2). For comparison, the inset shows the curve fits to the $\sqrt{H}$-dependence 
of the inverse $\kappa _{e}$ in Eq.(2). 
}
\label{KH}
\end{figure}

Next, we show in the Fig.\ref{KH} the magnetic field dependence of the in-plane thermal conductivity 
$\kappa _{ab}(H)$ of single crystals  (Y$_{1-x}$,Sm$_{x}$)Ba$_{2}$Cu$_{3}$O$_{7-\delta}$ ($x$=0, 0.1 and 1.0) and YBa$_{2}$(Cu$_{1-y}$Zn$_{y}$)$_{3}$O$_{7-\delta}$ ( $y$=0.0071). 
The $\kappa _{ab}(H)$ data are normalized by the zero field value.
At low fields, the value of $\kappa _{ab}$ exhibits a rapid drop around  nearly $T_{c}$/2
and it then tends to saturate at high fields up to 14 T. 
The applied field causes a substantial decrease in $\kappa(H)/\kappa(0)$ by $\sim30\%$ at $x$=0, while
$\kappa(H)/\kappa(0)$ of Sm-substituted samples shows a smaller degradation from $\sim16\%$ at $x$=0.1 to $\sim12\%$ at $x$=1.0. 
In the mixed state of conventional superconductors, we believe that thermal carriers (quasiparticles and phonons) are scattered by the magnetic vortices, resulting in a substantial decrease in $\kappa$. 

Now, we try to fit the $\kappa _{ab}(H)$ data to  the electronic contribution model \cite{YU94,PO95}.
 The electronic thermal conductivity in magnetic fields, $\kappa_{e}(T,H)$ , is rewritten using both $\kappa_{e}(T,0)$and $\ell_{0}$ as follows;
\begin{equation}
\kappa_{e}(T,H)=\frac{1}{3}C_{e}(T)v_{f}\ell(T,H)=\kappa_{e}(T,0)\frac{\ell(T,H)}{\ell_{0}(T)}=
\frac{\kappa_{e}(T,0)}{1+\ell_{0}(T)/ \ell_{v}(H)}
\end{equation}

%  $\kappa_{e}(T,H)=C_{e}(T)v_{f}\ell(T,H)/3$
% $=\kappa_{e}(T,0)$$\ell(T,H)/ \ell_{0}(T)$ 
% $= \kappa_{e}(T,0)$$/(1+\ell_{0}(T)/ \ell_{v}(H))$ . 
Here, the total mean free path of quasiparticles is expressed such as $\ell(T,H)^{-1}=\ell_{0}(T)^{-1}+\ell_{v}(H)^{-1}$,
where $\ell_{0}$ and $\ell_{v}$ are the mean free path of quasiparticles in zero field and in sufficiently
strong fields, respectively. 

Accordingly, the field dependence  of the  thermal conductivity ,$\kappa(T,H)$,is fitted using the following formula 
\begin{equation}
\kappa(T,H)= \kappa_{e}(T,H)+ \kappa_{ph}(T)=\frac{\kappa_{e}(T,0)}{1+p(T)H^{n}}+\kappa_{ph}(T) 
\label{fit}
\end{equation}
,where we set  $p(T)$ as $ \sigma \ell_{0}/ \phi_{0}$ and  $\ell_{0} /k \sqrt{\phi_{0}}$,  for $n$=1 and 0.5, respectively
($\phi_{0}$, the quantum flux).
%It is noted that the coefficient $p(T)$ in the denominator is proportional to $\ell_{0}$.
In the case of $n=1$, $\sigma$ represents  a scattering cross section per vortex cores in transport \cite{CL68}.
When the quasiparticle scattering from vortex cores is proportional to the number of the magnetic quantum flux, we get  $\ell_{v}(H)^{-1}$=$\sigma H/\phi_{0}$\cite{ON99}. 
Thus, for the exponent $n$=1 in eq. (\ref{fit}), we take into account for  the mean free path of quasiparticles $\ell_{v}(H)$
from the scattering due to vortex cores,  
 in addition to $\ell_{0}$ for scattering processes in the Meissner state.  
 Using the $\sigma$ = 90 $\AA $ in YBCO \cite{ON99}, $\ell_{v}(H)$ is $\sim 2.3\times10^3 H^{-1}$  \AA T$^{-1}$.  
On the other hand, in the case of $n$=0.5,  assuming the quasiparticle scattering due to screening currents around 
vortex cores in a disordered vortex array,
the vortex contribution $\ell_{v}(H)$ is proportional to the average lattice constant of a vortex lattice
$a_{v}$ i.e., $\ell_{v}(H)=ka_{v}\simeq k\sqrt{\phi_{0}/H}$, where $k$ is a constant independent of fields
\cite{FR99}. Using $k$=0.5 in \cite{FR99}, $\ell_{v}(H)$ becomes  $\sim 2.2\times10^2H^{-1/2}$  \AA T$^{-1/2}$.

Our reliable fits of the $\kappa(H)$ data to Eq.(\ref{fit}) give the two parameters,  
$\kappa_{e}(T,0)/ \kappa$(=1-$\kappa_{ph}/ \kappa$)  and $p(\propto \ell_{0})$, at selected temperatures 
as summarized in Table 1.  It is truly  that our fitting procedures yield a better fit for the $n=1$ than that for the $n=0.5$
, but it is difficult to distinguish a $H$-linear or $\sqrt{H}$-dependence in the denominator of $\kappa(H)$ in Fig.\ref{KH}.
In $H$-linear fits, we notice the underestimation in electron component for all crystals if we accept the phonon component
estimated from the as-grown data. For example, in the pure crystal of $x$=0,  the $H$-linear fit yields $\kappa_{e}\sim32\%$, while 
in the $\sqrt{H}$ fit  we get $\kappa_{e}\sim50\%$.  
This tendency in $\kappa_{e}$ below $T_{c}$ appears in previous works on the field dependences of $\kappa(H)$  for underdoped YBa$_{2}$Cu$_{3}$O$_{6.63}$ 
and Tl$_{2}$Ba$_{2}$CuO$_{6}$\cite{YU94,ON99}. 
Such the underestimation in $\kappa_{e}$ will enable us to consider a significant modification of the free-electron Lorenz number  in the WF law for the strongly electron correlated system such as high- $T_{c}$ cuprates, giving the reduced electron thermal conductivity \cite{ZH00}. 

Finally, let us examine the fitting parameter $p$ listed in Table\ref{table1}. From $p\propto \ell_{0}$, 
we obtain  the mean free path of quasiparticles $\ell_{0}\sim1.1\times10^{-7}$m for $n$=1 at the pure crystal. 
Here, we used the cross section of $\sigma$ = 90 $\AA $ in YBCO \cite{ON99}. 
In a similar way, for the $n$=0.5, $\ell_{0}\sim7.8\times10^{-6}$m with $k$=0.5 in \cite{FR99}.
Both the former and latter fits give $\ell_{0}\sim$1000$\AA$ at 40K for pure $x$=0 crystal. 
From Table\ref{table1},   
we guess that  $\ell_{0}$ at the Sm substituted crystal  is about one-third as long as the value of $\ell_{0}$  of the pure crystal, whether we use the $H$-linear fit or  the $\sqrt{H}$. 
If the normal state $\kappa_{e}(T)$ is proportional to  $\ell_{0}(T)$ at $T\geq T_{c}$ , then the  $\kappa_{e}^{n}$ plots in Fig.\ref{KE}(a)give the ratio of  $\ell_{0}$ at $x$=0.1 to that at $x$=0 $\sim0.26$  at higher $T$ , 
This value is not far from the above results in the  $\ell_{0}$ ratio ($\sim1/3$) around a half of $T_{c}$. 
Moreover, for the lightly Zn-doped crystal, our fits to the data indicate the reduced mean free path of quasiparticles. 
 A pair-breaking effect due to Zn-doping makes quasiparticles increase, giving the strongly electron-electron scattering enhanced below $T_{c}$. This finding has a close relationship with the reduced peak observed in $\kappa_{ab}$ of YBa$_{2}$(Cu$_{1-y}$Zn$_{y}$)$_{3}$O$_{7-\delta}$ with y=0.0071.  

\section{Summary}

We have studied  the  in-plane thermal conductivity of large single crystals (Y$_{1-x}$,Sm$_{x}$)Ba$_{2}$Cu$_{3}$O$_{7-\delta}$ ($x$=0, 0.1, 0.2 and 1.0) and YBa$_{2}$(Cu$_{1-y}$Zn$_{y}$)$_{3}$O$_{7-\delta}$ ( y=0.0071) 
as  functions of temperature and magnetic field. 
The nonlinear field profile of $\kappa(H)$ observed around a maximum peak is discussed on the basis of the electron contribution model. We tried to make the fits of the $\kappa(H)$ data to the $H$-linear or $\sqrt{H}$ dependence of the inverse electronic thermal conductivity in Eq.(2). 
The local distortion due to the Sm substitution for Y site suppresses both the phonon and electronic contributions, accompanied by the substantial decrease in   $\ell_{0}$  at 40K in comparison with that of the $x$=0 pure sample. 
The light Zn doping for CuO$_{2}$ planes 
does not contribute the phonon conduction but affects solely the electron component below $T_{c}$, resulting in the reduced mean free path of quasiparticles in the superconducting state.
For SmBCO, a suppressed maximum in  $\kappa(T)$ arises from the partial substitution of Sm for Ba site  
,which is closely related to the slight degradation in $T_{c}$. 

One of authors (X Yao) would like to thank NBRPCE(No.2006CB601003) and SSTC (No.055207077) for financial support.
We are grateful to F.Sato and Prof.S.Kambe for the iodometric analysis.  

\begin{table}[p]
\caption{\label{table1}  The fitting parameters $\kappa_{e}/\kappa$  and $p$
from the fit of $\kappa _{ab}(H)$ data to eq.(\ref{fit})for single crystals (Y$_{1-x}$,Sm$_{x}$)Ba$_{2}$Cu$_{3}$O$_{7-\delta}$ ($x$=0, 0.1 and 1.0) and 
YBa$_{2}$(Cu$_{1-y}$Zn$_{y}$)$_{ 3}$ O$_{7-\delta}$ ($y$=0.0071).
For comparison, we show the electronic component near $T_{c}$/2, estimated from  the $\kappa$ data of as-grown crystals described in the text. On the last column, the electronic components in the normal state are given using the WF law.  }

\begin{center}
\begin{tabular}{cccccccccc} \hline\hline
Sample& &&$n=1$&&&$n=0.5$&&estimation&WF law\\
Composition &$T$ &  $\kappa_{e}/\kappa$ &$p$&$\ell_{0}$& $\kappa_{e}/\kappa$ &  $p$&$\ell_{0}$&$\kappa_{e}/\kappa$& $\kappa_{e}/\kappa$(150K)\\
$x$&(K)&(\%)& &\AA&(\%)&&\AA& (\%) &(\%)\\
\hline
0.0& $40 $ & 32& 0.46 &$1.1\times 10^3$ & 50&0.34&$7.8\times 10^2$& 57&31 \\
0.1& $40 $ & 22 & 0.17 &$4.1\times 10^2$  &53 &0.11& $2.5\times 10^2$ &45&12\\

1.0& $30 $ & 14& 0.26& $6.2\times 10^2$ &35&0.13&$3.0\times 10^2$ && 10 \\

$y$=0.007 &  $40$ & 11& 0.17&$4.1\times 10^2$ & 30&0.09&$2.1\times 10^2$ &29&22  \\\hline\hline\\

\end{tabular}
\end{center}
%\end{ruledtabular}
\end{table}

\end{document}